\documentstyle[preprint,pra,aps,epsfig]{revtex}
\tighten
\begin{document}
\preprint{}
\draft

\def\lsim{\mathrel{\rlap{\lower4pt\hbox{\hskip1pt$\sim$}}
    \raise1pt\hbox{$<$}}}         %less than or approx. symbol
\def\gsim{\mathrel{\rlap{\lower4pt\hbox{\hskip1pt$\sim$}}
    \raise1pt\hbox{$>$}}}         %greater than or approx. symbol
\def\dblint{\mathop{\rlap{\hbox{$\displaystyle\!\int\!\!\!\!\!\int$}}
    \hbox{$\bigcirc$}}}
\def\ut#1{$\underline{\smash{\vphantom{y}\hbox{#1}}}$}
\def\Pom{{\bf I\!P  }}
\def\be{\begin{equation}}
\def\ee{\end{equation}}
\def\bq{\begin{eqnarray}}
\def\eq{\end{eqnarray}}
\def\bm{\boldmath}
 
%%%%%%%%%%%%%%%%%%%%%%%%%%%%%%%%%%%
%       Definitions for boldmath

\mathchardef\alpha="710B
\mathchardef\beta="710C
\mathchardef\gamma="710D
\mathchardef\delta="710E
\mathchardef\epsilon="710F
\mathchardef\zeta="7110
\mathchardef\eta="7111
\mathchardef\theta="7112
\mathchardef\iota="7113
\mathchardef\kappa="7114
\mathchardef\lambda="7115
\mathchardef\mu="7116
\mathchardef\nu="7117
\mathchardef\xi="7118
\mathchardef\pi="7119
\mathchardef\rho="711A
\mathchardef\sigma="711B
\mathchardef\tau="711C
\mathchardef\upsilon="711D
\mathchardef\phi="711E
\mathchardef\chi="711F
\mathchardef\psi="7120
\mathchardef\omega="7121
\mathchardef\varepsilon="7122
\mathchardef\vartheta="7123
\mathchardef\varpi="7124
\mathchardef\varrho="7125
\mathchardef\varsigma="7126
\mathchardef\varphi="7127
\mathchardef\nabla="7272
 
\font\dozeb=cmmib10 scaled \magstep1
\font\dozesyb=cmbsy10 scaled \magstep1
\font\dezb=cmmib10
\textfont9=\dozeb
\scriptfont9=\dezb
\def\bm{\fam9}
\textfont10=\dozesyb

%
%%%%%%%%%%%%%%%%%%%%%%%%%%%%%%%%% TITLE PAGE
%
\title{Quark Matter\\   
in the Chiral Color Dielectric Model}
\author{Alessandro Drago$^{a}$, Manuel Fiolhais$^{b}$ and Ubaldo Tambini${^a}$}
\address{$^{a}$Dipartimento di Fisica, Universit\`a di Ferrara, 
and INFN, Sezione di Ferrara,\\
Via Paradiso 12, Ferrara, Italy 44100\\
$^{b}$Departamento de F{\'i}sica da Universidade \\
and Centro de Fisica Te{\'o}rica, P-3000 Coimbra, Portugal}
\date{\today}
\maketitle
%
%%%%%%%%%%%%%%%%%%%%%%%%%%%%%%%%% ABSTRACT
%
\begin{abstract}
We study the equation of state (EOS) of quark matter at zero temperature, 
using the Color
Dielectric Model (CDM) to describe confinement.
Sensible results are obtained in the version of the CDM for which confinement
is imposed {\it smoothly}. The two--phases version of the model
turns out to give unrealistic results for the EOS. 
Chiral symmetry plays a marginal r\^ole
and the quarks are massive till high densities. The deconfinement
phase transition is smooth and unlikely to be first order.
Instabilities of the quark matter and the gap equation are discussed.
\end{abstract}
%
%%%%%%%%%%%%%%%%%%%%%%%%%%%%%%%%% PACS NUMBERS
%
%
%%%%%%%%%%%%%%%%%%%%%%%%%%%%%%%%% PAPER CONTENT
%
\narrowtext
\section{Introduction}

The study of the Equation Of State (EOS) of Quark Matter (QM)
has become a fashionable topic in view of the next experiments
using heavy ions in program at RHIC(BNL) and at LHC(CERN) \cite{qm93}. 
Furthermore
the inner structure of neutron stars is now under investigation:
the connection between the composition of the star and the
cooling time, which can be measured, allows to discriminate among
the various models, indicating the possible existence of a quark
matter phase (see, e.g. \cite{stars}).
The study of the equation of state of matter at high densities
can also give usefull informations to traditional nuclear physics, since one can
search heavy nuclei for precursor phenomena, both of the deconfinement
and/or of the chiral restoration phase transition (for a review see
\cite{brown}).

In many model calculations of the deconfinement phase transition
the frame of the MIT bag model has been used\cite{rafe,giapu}. 
In such a way a first order
deconfinement phase transition is obtained (apart from specific, {\it
ad hoc} choices of the model parameters) and the deconfinement phase
transition coincide with the chiral restauration one. At densities
and temperatures slightly bigger than the critical ones the right
degrees of freedom are already quarks, having current masses, and perturbative
gluons. There are anyway indications, from lattice calculations, that
at temperatures bigger than the critical one non perturbative effects
are still present in the quark--gluon plasma \cite {blaizot}.

In this paper we study the EOS of QM
using the Color Dielectric Model (CDM) to describe confinement 
\cite{pirner}. We shortly review CDM in sect. 2.
This model has been widely used to study both the static 
and the dynamical properties of the nucleon. Morover
it can be used to describe many--nucleon systems: for
a two nucleon system it allows to compute a nucleon--nucleon potential
qualitatively similar to the ones used in nuclear physics \cite{seiquark};
in the case of a homogeneous, infinite system of nucleons 
the CDM can be used to construct a nonlinear version of the
Walecka model \cite{birsewal}.

The aim of our work is to extend previous calculations
of the deconfinement phase transition, 
where the same model has been used \cite{mitia}.
An important point of our calculation
will be to fix the model parameters in order to reproduce the basic static
properties of the single nucleon, 
as was already done in the study of the nucleon
structure functions \cite{structure}. 
We will later use the same parameters to study the
EOS of QM, which we define as a system of totally
deconfined quarks. In such a way the study of the 
QM's EOS will turn out to be a severe test for the different versions of
the CDM, and we will be able to make some predictions of the properties
of matter at high densities.
%and later to study the stability of the quark matter
%and the Bethe--Goldstone gap equation, to shed some light on the details
%of the deconfinement transition. 
Within CDM (with a double minimum potential
for the scalar field), we will investigate the possibility
of getting a {\it scenario} similar to the one 
described by the MIT bag model, with two phases undergoing a sharp
first order phase transition (sect. 3). Our results show that such a description
is incompatible with the CDM. We than study another version of the CDM 
(with a single minimum potential for the scalar field),
where confinement is imposed more smoothly, and we get 
a sensible EOS for QM, without a sharp deconfinement transition.

An important feature of this deconfined quark matter is that quark's
masses are big till high densities. Chiral restauration and deconfinement
do not occur at the same density. The reasons why chiral simmetry is
restaured so slowly are discussed in sect. 4.

In the last sections we analyze the properties of QM, as described
in the CDM.
In sect. 5 the stability of quark matter is studied, using the technology
of the response function. In sect. 6 we consider a 
%Bethe--Godstone
gap equation, trying to 
%relate the instabilities found in the previous
%section to 
understand 
the formation of quarks' clusters. Finally sect. 7 is devoted
to the concluding remarks.

\section{The Color Dielectric Model}
\subsection{The model lagrangian}
In this section we shortly summarize the main features of the CDM.
For a comprehensive review see the ones by Pirner \cite{pirner}
and by Birse \cite{birserev}. 

We will use a chiral invariant version of the CDM,
as the one used in Ref.\cite{mitia} and also in Ref.\cite{structure}.
The Lagrangian reads
%\cite{NF93,Oro,Bro87}
\begin{eqnarray}
{\cal L} &=& i\bar \psi \gamma^{\mu}\partial_{\mu} \psi
       +{g\over \chi} \, \bar \psi\left(\sigma
+i\gamma_5\vec\tau\cdot\vec\pi\right) \psi        \nonumber
 \\
         &+&{1\over 2}{\left(\partial_\mu\chi\right)}^2
                       -U \left(\chi\right)
                       +{1\over 2}{\left(\partial_\mu\sigma \right)}^2
                       +{1\over 2}{\left(\partial_\mu\vec\pi\right)}^2
                       -U\left(\sigma ,\vec\pi\right)   \, ,
\label{eq:in1}
\end{eqnarray}
where $U(\sigma ,\vec\pi)$ is the usual mexican-hat potential, 
 as in ref. \cite{NF93}.
${\cal L}$  describes a system of interacting quarks, pions, sigmas and
a scalar-isoscalar chiral singlet field $\chi$, 
whose potential $U(\chi)$ has an absolute minimum for $\chi =0$.
In such a way, in the case of the single nucleon problem,
the quarks' effective mass $-g\sigma/\chi$ diverges
outside the nucleon. The simplest potential for $\chi$ is a quadratic one:
\be
U(\chi)=\frac{1}{2} M^2 \chi^2.
\label{eq:in3}
\ee
Using this potential, $\chi$ fluctuates around zero, reaching
this value asymptotically, at large distances. In the following we
will refer to this version of the model as the Single Minimum (SM) one.
Another possibility is to use a potential for $\chi$ having a Double
Minimum (DM): 
%is given by
\be
U(\chi)=\frac{1}{2} M^2 \chi^2 \left[ 1 + \left(\frac{8
\eta^4}{\gamma^2} - 2 \right) \frac{\chi}{\gamma M} + \left(1-\frac{6
\eta^4}{\gamma^2} \right) \, \left( \frac{\chi}{\gamma M} \right) ^2
\right],
\label{eq:in2}
\ee
where the absolute minimum is still in the origin and the relative one is
in $\chi=\gamma M$. Choosing appropriately the parameters it is possible
to obtain solutions for the single nucleon where the $\chi$ field
interpolates between the relative minimum at the center of the
nucleon and the absolute minimum at big distances. In the following,
we will consider for DM only solutions where $\chi$ has the above discussed
behaviour. A complete analysis of the various possible solutions of the
model (in a non--chiral--invariant version) can be found in Ref.\cite{maryland}.

\begin{figure}
\centerline{\hbox{
\psfig{figure=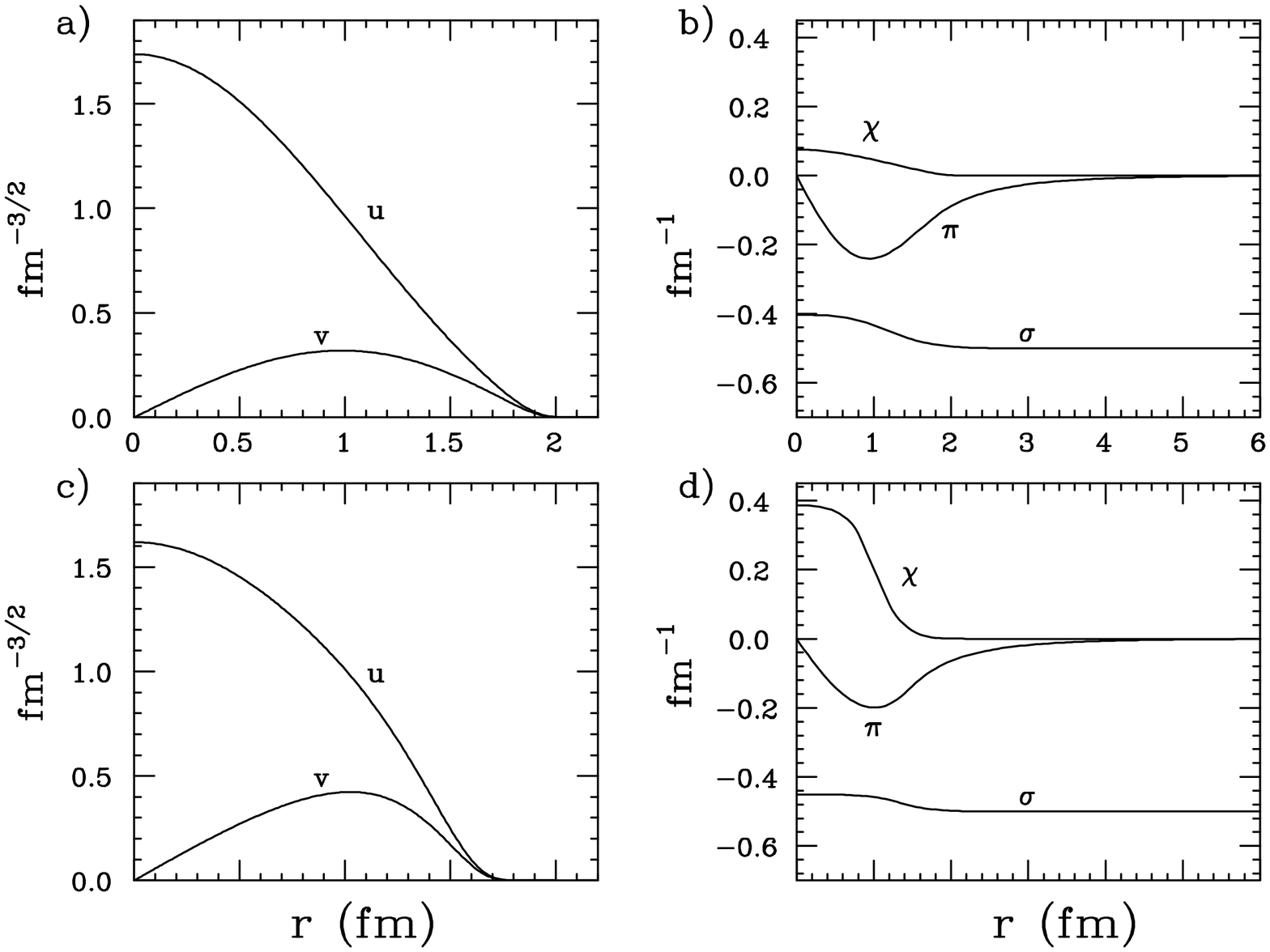,angle=90,width=14cm}
}}
\caption{Typical solutions of the model for single nucleon problem in
the SM (a,b) and DM (c,d) versions.}
\label{fig:FIG1}
\end{figure}

In Fig.\ref{fig:FIG1} we show two typical solutions of the model for the single
nucleon problem. As it appears, in the DM case the transition between
the interior of the nucleon and the external region is sharper than
in the SM, where all the fields have a very smooth behaviour.
Correspondingly, the kinetic energy contribution (which comes mainly
from the quarks) will be bigger in DM than in SM. This point will be
relevant when studying the EOS of QM.

The Lagrangian is chiral invariant, and it can be considered a 
confining version of the traditional $\sigma$--model.
An important point concern the value of the chiral fields, the pion
and the sigma, in this model, in the 
single nucleon case. These fields are always near their vacuum value.
This point has been checked out numerically several times and some
euristic explanations have been proposed \cite{birserev,chirale}.
%\footnote{A simple argument is the following: in order to have confinement
%the quark effective mass $-g\sigma/\chi$ has to become large and
%{\it positive} at big distances. If some large fluctuation in the
%chiral field is allowed, there would be a region inside the
%nucleon where the effective quark mass would change the sign (the 
%$\chi$ field has to have the same sign everywhere, not to cross zero).
%If this happens, the so find solution of the field's equations is
%instable: another solution with a lower energy can be obtained just
%reducing the value of $\chi$ in the region where the effective quark mass
%is negative.}
As a consequence, the chiral fields cannot
`wind' around the mexican hat potential. The pion is thus just a perturbation
and cannot develop a non--trivial topology. 
%We will come back to this
%point when discussing chiral symmetry in QM.
%

\subsection{Fixing the parameters}

The parameters of the model are: the chiral meson masses
$m_\pi=0.14$ GeV, $m_\sigma=1.2$ GeV, the pion decay constant
$f_\pi=0.093$ GeV, the coupling constant $g$, and the parameters
appearing
in the quadratic (\ref{eq:in3}) or in quartic $\chi-$potential (\ref{eq:in2}). 

The free parameters are fixed to reproduce the basic properties of the
nucleon.
% the mass and the radius. 
%after having solved the mean fields equations two projections
%have to be performed: the projection on a zero momentum state of the 
%total nucleon and a projection on the right spin and isospin quantum
%numbers
%The last projection is essential, because the mean field equation
%are solved using the hedhehog ansatz, 
At the mean field level we use the hedgehog ansatz which is an eigenstate of the
so called Grand Spin $\vec G=\vec S + \vec I$, and is a superposition
of various bare nucleon and delta states.

In order to describe the single nucleon state we performed a 
double projection on
linear and angular momentum eigenstates from the hedgehog, whose details
can be found in Ref.\cite{NF93}. 

For the DM version of the model we use the sets of parameters of Ref.
\cite{physlett}, for which a good description of the static properties
of the nucleon was obtained.

For the SM version such a set of parameters was not available in the
literature. We fixed the parameters $g$ and $M$ to reproduce the
experimental value of the average mass of the nucleon and of the delta,
and the isoscalar radius of the nucleon. Choosing $g=0.02$ GeV and
$M=1.7$ GeV we got: $(E_N+E_\Delta)/2=1.112$ GeV (exp.val.$=1.085$GeV) and 
$<r_N^2>^{1/2}_{isoscalar}=0.82$ fm (exp.val.$=0.79$ fm). 
These values depend essentially only on the quantity $G=\sqrt{gM}$. A detalied
presentation of the single--nucleon properties in SM version of the CDM
will be presented elsewhere \cite{progre}.

To perform an exhaustive analysis of the various versions of the CDM, we
considered also the possibility of having an effective mass term for
the quarks in which the $\chi$ field appears with a power different
from one: $m_q=-g^p\sigma/\chi^p$. Studying this possibility in the
SM version, and for $p=2$, 
a good description of the single nucleon properties can
be achieved using $g=0.02$ GeV and $M=1.10$ GeV ($(E_N+E_\Delta)/2=1.102$GeV
and $<r_N^2>^{1/2}_{isoscalar}=0.78$fm).

Several important differences exist between the CDM and the
MIT bag model. In the latter model, inside the bag the quarks have
current masses of few MeV. The bag is stabilized through the introduction
of a big {\it vacuum pressure}, of the order of $150 {\rm MeV/fm}^3$. 
Perturbative
gluons are considered to be the right degrees of freedom inside the bag.
In the CDM model, in all versions, 
the effective quark mass is everywhere bigger than 
a number of the order of $100$MeV,
hence chiral symmetry is broken and Goldstone bosons are the right
degrees of freedom. We will come back later to this point, comparing
the EOS of QM as computed in the chiral CDM with the one computed
in a non--chiral version \cite{gluoni}. In the CDM a {\it vacuum pressure}
is also present, coming from the $\chi$ field: this pressure is
roughly constant inside the nucleon in the DM case, where it equals 
$U(\chi=\gamma M)=M^4 \eta ^4$,
whilst in the SM case the pressure ${1\over 2}M^2 \chi (r)^2$ depends
on $r$. It is important to stress that in the DM version of the CDM model
the pressure is very small, of the order of few MeV, and this point
will also be important when discussing the EOS of QM.

\section{MEAN FIELD APPROXIMATION TO THE EOS OF QM}

In this section we will study the EOS of QM, using the Lagrangian of the 
chiral CDM (1--3).

For QM we mean a system of totally deconfined quarks,
described using plane waves, with the $\chi$,
pion and sigma fields having a constant value, given by the Euler--Lagrange
equations.

The total energy of QM in the mean field approximation is the following:
\be
E_{QM}=12 \, V \int {d{\bf k} \over (2\pi)^3}
\sqrt{{\bf k}^2 + ({g \bar\sigma \over\bar\chi})^2} \,
\theta (k_F-k) + V U(\bar \chi) + V U(\bar \sigma,\vec\pi=0),
\label{eq:in4}
\ee
where $k_{F}$ is the Fermi momentum of quarks, 
$\bar \chi$ and $\bar\sigma$ are the solutions of the coupled equations
\be
{dU(\chi) \over {d \chi}} \Bigm|_{\chi =\bar \chi}=
-g\bar\sigma {\rho _S (\bar \chi,\bar\sigma)\over \bar \chi^2},
\label{eq:in5}
\ee
\be
{dU(\sigma,\vec\pi=0) \over {d \sigma}} \Bigm|_{\sigma =\bar \sigma}=
-g {\rho _S (\bar \chi,\bar\sigma)\over \bar \chi},
\label{eq:in6}
\ee
and the scalar density $\rho _S (\bar \chi,\bar\sigma)$ is given by
\be
\rho _S(\bar \chi,\bar\sigma)=
<\overline\psi \psi>=12 \int {d {\bf k}\over (2\pi)^3}
{{g\bar\sigma/\bar \chi}\over \sqrt{{\bf k} ^2 + ({g\bar\sigma/ \bar \chi})^2}}
\theta (k_{F}-k).
\label{eq:in7}
\ee
In the mean field approximation, for an homogeneous infinite system,
$<\vec\pi>=0$, so the pionic field is not contributing (of course it
indirectly enters the EOS, because our model parameters are fixed in the
single nucleon problem, where $<\vec\pi>\ne 0$). This is drawback of the
mean field approximation, when applied to homogeneous infinite systems.
A way to circumvent this problem is discussed by Ghosh and 
Phatak \cite{phatakpion}.

It is important to analyze the behaviour of the $\chi$ field,
both in the SM and in the DM version of the model,
when the CDM is used to describe a collection of nucleons at increasing
densities. First of all, in the DM model a critical density
exist, for which the $\chi$ field undergo a discontinous jump.
In the appendix a proof of this statement is given, based on the 
study of a two--nucleon system for various internucleon distances.
For distances smaller than a critical one, the $\chi$ field,
in the region between the two nucleon, will cease to
interpolate between the two minima, as discussed in Sec.2B, and will
stay near the relative minimum. This transition cannot be made continuous.
In the DM version of the CDM, the deconfinement phase transition is therefore
a discontinous first order transition.
In the SM version of the model nothing similar can happen. Of course this
is not enaugh to conclude that, in this case, 
the deconfinement phase transition is not first order. We will study
more in detail what happens in Sec.5 and 6.

\begin{figure}
\centerline{\hbox{
\psfig{figure=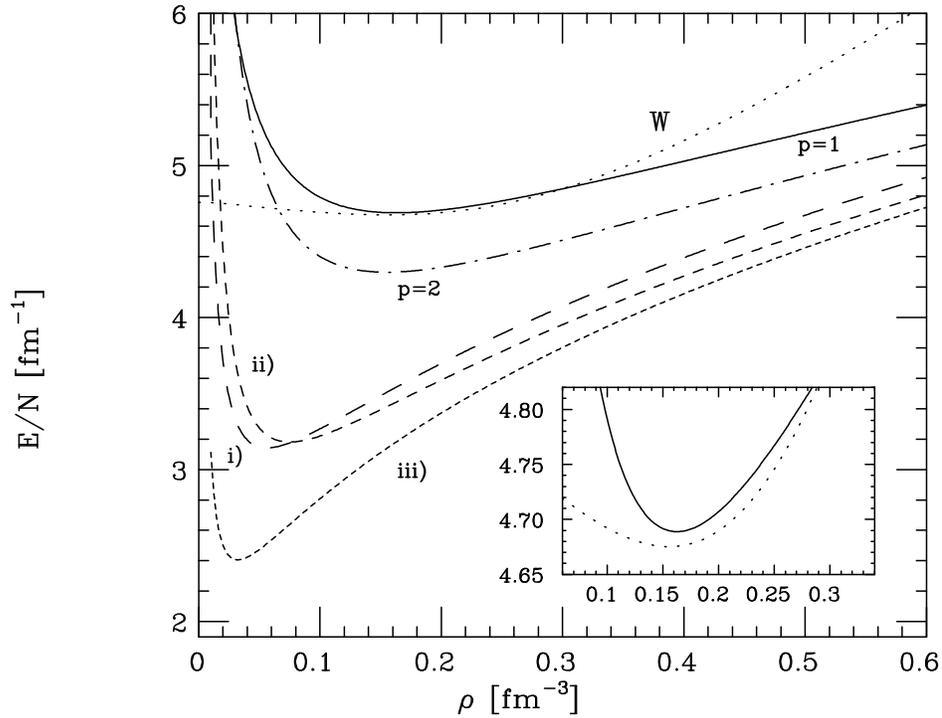,angle=90,width=15cm}
}}
\caption{EOS of QM and EOS of nuclear matter in the Walecka model (dotted
line).We show the EOS of QM in CDM for the DM version of the model (dashed
lines i) ii) iii) ) and for the SM version with p=1 (solid line) and p=2
(dot-dashed line)
.}
\label{fig:EOS}
\end{figure}

We will now compare the EOS of QM, 
as computed in our model, with the EOS of nuclear matter as 
obtained in the Walecka model \cite{walecka}.
In Fig.\ref{fig:EOS} our results for the EOS are shown. 

In the DM case we used the parameter sets\cite{NF93,physlett}:
\begin{itemize}
\item[ i):]$g=0.059~$ GeV, $M=1.4~$GeV,$\gamma=0.04$,$\eta=0.06$
\item[ ii):]$g=0.029~$ GeV, $M=1.2~$GeV,$\gamma=0.06$,$\eta=0.06$
\item[ iii):]$g=0.0235~$ GeV, $M=1.6~$GeV,$\gamma=0.03$,$\eta=0.06$
\end{itemize}
As it appears, the
energy per baryon number in DM is very small, its value being below the
one given by the Walecka model for almost all densities. The DM version
of the CDM is therefore unrealistic when used to describe the EOS
of QM. A similar result was obtained in a calculation performed in a
non--chiral version of CDM \cite{gluoni}. 
The reasons for such a {\it d\'eb\^acle} are to be find in the very small
value of the pressure in DM. If one would add `by hand' a pressure's
contribution of the order of $150 MeV/\rho$ to the
energy per baryon shown in Fig.\ref{fig:EOS} for DM, 
one will get a result similar to the one obtained
using the MIT bag model. The problem is that such a big pressure cannot
be obtained in the CDM, because corresponds to a solution for the single
nucleon problem where the quark fields are very steep: as it has been
shown by Leech and Birse \cite{birsecm}, the center of mass motion cannot
be projected out consistently in this case, 
and the mean field approximation is no more
a good starting point.

In the SM with $p=1$ case (see Sect.2B), on the other hand, 
the mean field approximation to the
EOS of QM gives a sensible result: the energy per baryon number is
bigger in the QM phase than in the hadronic phase for all densities
smaller or of the order of $\rho _{eq} =0.17N/fm^3$, the equilibrium 
density of nuclear matter. After this density the equation of state
of QM and the one of nuclear matter seems almost equivalent, till
densities of the order of $2 \rho _{eq}$, after which the energy of QM
is smaller than the energy of nuclear matter. We would like to remind
that we have not modified the parameters of the model, but we are 
sticking to the ones fixed to the static properties of the nucleon, as
discussed in Sec.2.

The last possibility we have considered is SM with $p=2$. Also in this
case, the EOS of quark matter that one obtains after fixing the
parameters is too low, as it can be seen from Fig.\ref{fig:EOS}. 

We can conclude from analysis of all the versions of the CDM, that
the most realistic EOS of the quark matter is obtained using the
model in which confinement is imposed in the smoothest way, i.e.
SM with $p=1$. In the following sections, only this
version will be considered.

The most relevant feature of the result for SM $p=1$, is the wide range of
densities for which the EOS of traditional nuclear matter and
the EOS of quark matter are almost equivalent. The difference
in energy between the two phases is of some tens' MeV, only.
It is remarkable that this `almost equivalence' starts at a density
of the order of $\rho_{eq}$. A natural interpretation of this 
result is that in the case of heavy nuclei, some precursor phenomena
of deconfinement could be seen (as swelling, for instance), but
no dramatic change is going to happen in the system till much
higher densities. In Sec.6 we will analyze the possibility that
the small energy gap between the two phases can be explained
taking into account correlations in QM, using Bethe--Goldstone equation.

\section{quarks' mass and chiral symmetry restauration}

We discuss now more in details the dependence of quarks' effective 
mass $m_q=-g\sigma/\chi$ on the density.
In this model two different mechanisms are at work to reduce $m_q$: 
one comes from the chiral field sigma, that 
moves from its vacuum value $-f_\pi$ as the density increases (in the
following we will not take into account a chiral--symmetry breaking
term giving mass to the pion, because it is irrelevant to what we want
to discuss).
The other mechanism that modifies $m_q$ is confinement,
through the $\chi$ field, which moves away from zero. 
As we will see, this second mechanism is the
relevant one till high densities. 

We restrict our discussion to the
SM $p=1$ version of the model, the one which gives sensible results for the
EOS of QM. In this case, eq.(5) can be {\it formally} solved, giving
\be
\bar\chi=(-{g\over M^2 }\bar\sigma<\bar\psi\psi>)^{1/3}.
\label{eq:in8}
\ee
This is not really the solution of eq.(5), because the scalar
density $\rho_S=<\bar\psi\psi>$ still depends on $\bar\chi$ through
the fermions' mass. It can nevertheless be used as 
an approximation to the real solution, if one neglects the difference
between $\rho=<\psi^{\dagger}\psi>$ and $\rho_S$ (this approximation
is not too bad, because, as we shall see, quarks' masses are not decreasing
very fast).

The `Maxican hat' potential, entering the Lagrangian (1), 
and parametrized as in Ref.\cite{NF93}, is
\be
U(\sigma,\pi)={m_\sigma^2\over 8 f_\pi^2}(\sigma^2 + \pi^2 - f_\pi^2)^2,
\label{eq:in9}
\ee
where $m_\sigma$ is the mass of the sigma field.
Using the formal solution for $\chi$ given in eq.(8), one can compute
the shift $\Delta$ of the sigma field from its vacuum value
\be
\bar\sigma=-f_\pi+\Delta
\label{eq:in10}
\ee
\be
\Delta={1\over m_\sigma^2}\left [{(gM)^2\over f_\pi}\right]^{1/3}
(\bar\psi\psi)^{2/3}\equiv C (\bar\psi\psi)^{2/3}.
\label{eq:in11}
\ee

Substituting in eq.(11) typical numbers, and in particular the ones
we used in Sec.3, the dimensional coefficient $C$ 
in eq.(11) turns out to be small, of the order
of 0.1 fm, and $\Delta$ is therefore also small, if compared to $f_\pi$, till
very high densities.

\begin{figure}
\centerline{\hbox{
\psfig{figure=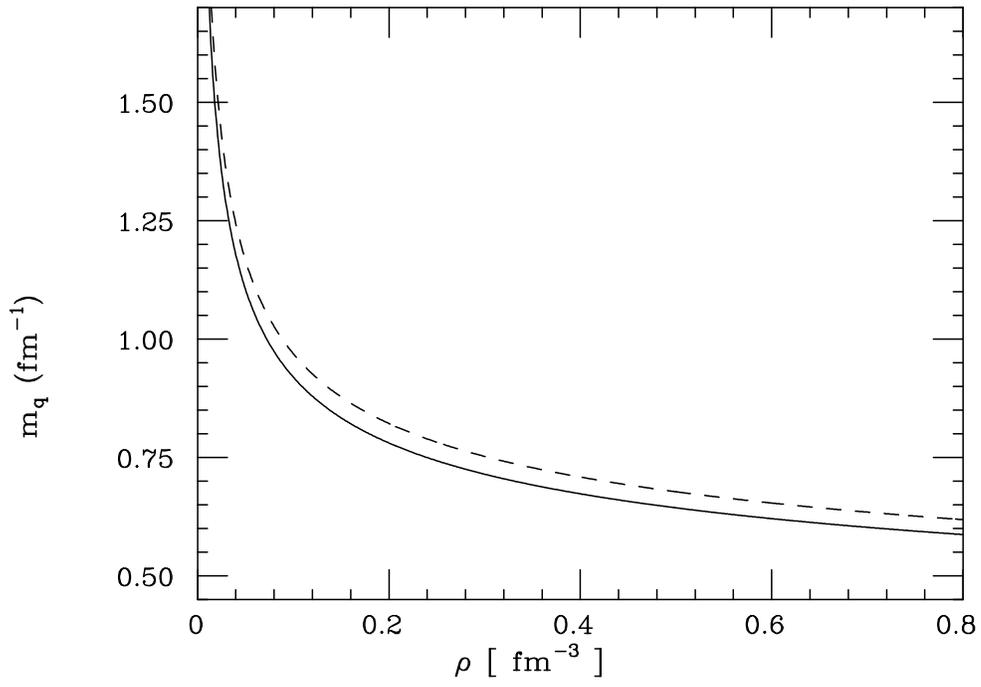,angle=90,width=15cm}
}}
\caption{Effective quark mass versus density: ....
.}
\label{fig:MASS}
\end{figure}

In Fig.\ref{fig:MASS} we compare the effective quark mass, as computed taking into
account both the reductions coming from the $\chi$ field and from
the $\sigma$ field (solid line), with the effective mass when the
$\sigma$ field is kept fixed at $-f_\pi$ (dashed line).
At a density of order of $3 \rho_{eq}$, the effective 
mass of the quarks is reduced from its value at $\rho_{eq}$ by a factor
$\simeq 0.8$. Comparing the two lines of Fig.\ref{fig:MASS}, one can see that
most of the effect is due to the confining $\chi$ field. 

\section{Quark Matter instabilities}
The more direct way to study the instabilities of a many--body system
is to look where its compressibility $K$ becomes infinite. The
compressibility is related to the pressure through the following
relations
\be
P=\rho^2 {\partial(E/N)\over \partial\rho}
\ee
\be
K^{-1}=\rho {\partial P\over \partial\rho}.
\ee
Here $(E/N)$ is the energy per particle and $\rho$ is the density.

\begin{figure}
\centerline{\hbox{
\psfig{figure=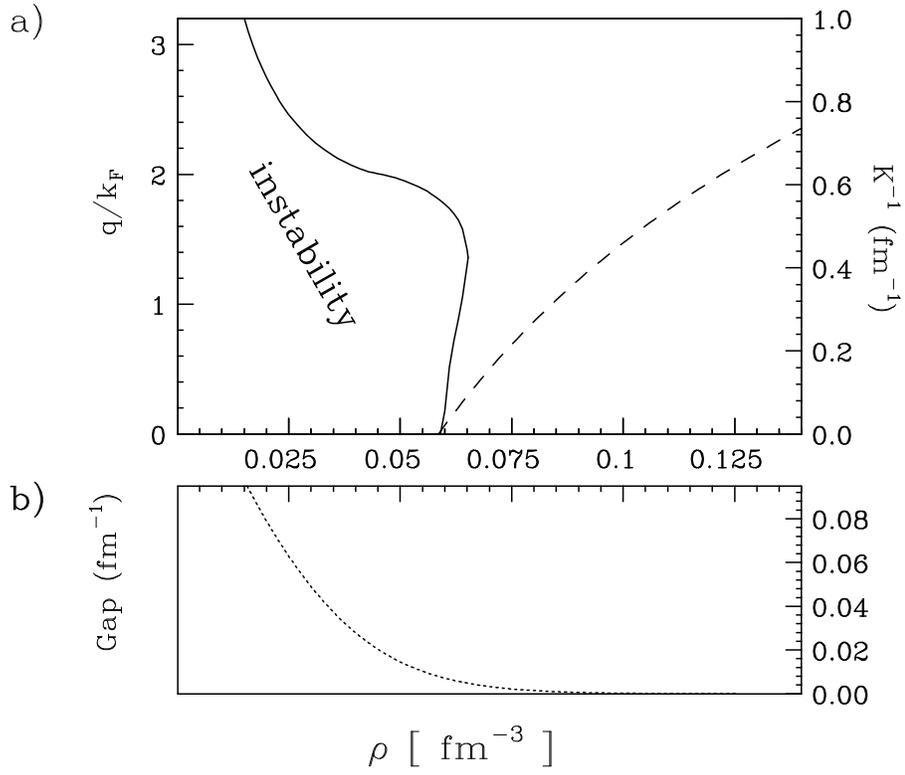,angle=90,width=15cm}
}}
\caption{Compressibility (dashed line), instability region (solid line),
and gap (dotted line) for the QM......
.}
\label{fig:INST}
\end{figure}

In Fig.\ref{fig:INST} the compression modulus $K^{-1}$ is shown (here and in the
following we will refer to the SM version of the CDM, only). 
As it can be seen, quark matter becomes unstable, in our calculation,
at a density smaller than $\rho_{inst}\simeq\rho_{eq}/2$. 
This instability is not
due to the fact that in this region the energy per particle of quark
matter is higher than that of nuclear matter, but it is related to the
possibility of creating undamped density fluctuations in the system, 
without spending energy. 

To clarify even more this point, one can try to reproduce the same
result, studying the collective excitations of the system, which can be
found searching for the poles of the propagators of the scalar fields,
dressed by the particle--hole polarization propagator.
In the following we will only discuss the collective states coming from
the propagation of the $\chi$ and the sigma field, leaving the discussion
of the pion propagator, and the related phenomena, to a future study.

To define the propagators of the $\chi$ and the sigma, we expand these fields
around their mean field value:
\be
\sigma=\bar\sigma+\tilde\sigma
\ee
\be
\chi=\bar\chi+\tilde\chi.
\ee
The mass term reads thus ($<\vec\pi>=0$)
\be
g{\bar\psi\sigma\psi\over\chi}=
g{\bar\psi(\bar\sigma+\tilde\sigma)\psi\over(\bar\chi+\tilde\chi)}=
g{\bar\psi\bar\sigma\psi\over\bar\chi}+
g{\bar\psi\tilde\sigma\psi\over\bar\chi}-
g{\bar\psi\bar\sigma\psi\over\bar\chi^2}\tilde\chi+...
\label{linear}
\ee
where we have expanded the $\chi$ field till first order in the
fluctuation. Since we are looking for the instabilities of the
system, i.e. for the situations in which an (arbitrary small) fluctuation
around the mean field develops spontaneously and propagates undamped,
it is enough to consider a first order expansion in $\tilde\chi$.
%As we shall see, this is not a totally satisfactory expansion,
%since the instabilities, computed using the RPA formalism, will not
%start at exactly the same density we got studying the compressibility.
%This discrepancy, which is not present {\it e.g.} in the Walecka model 
%\cite{matsui}, comes from the approximate linearization of the lagrangian
%around the mean field, whereas the Walecka model's lagrangian is already
%linear.

From the linearized lagrangian (\ref{linear}), we read the couplings
between the scalar fields $\tilde\chi$ and $\tilde\sigma$ and
the quark fields
\be
g_{\tilde\chi}=-g{\bar\sigma\over\bar\chi^2}
\ee
\be
g_{\tilde\sigma}={g\over\bar\chi}.
\ee

The masses corresponding to the $\tilde\chi$ and to the $\tilde\sigma$
fluctuations are given by the following relations
\be
M^{*2}_{\tilde\chi}\equiv<{\partial^2 \over\partial\tilde\chi^2}
\left [U(\chi)-g{\bar\psi\sigma\psi\over\chi}\right ]>=
M^2-2g{\bar\psi\sigma\psi\over\chi^3}=3M^2
\label{mchi}
\ee
\be
m^{*2}_{\tilde\sigma}\equiv<{\partial^2 U(\sigma,\vec\pi)\over
\partial\tilde\sigma^2}>.
\ee
In the previous equations, the brakets $<>$ indicate mean value in the
mean field approximation, where the fluctuations are set equal to zero.
In eq.(\ref{mchi}), we have used the field equation for $\bar\chi$ 
(\ref{eq:in5}),
to eliminate the dependence on the fields.
The mass of the $\tilde\sigma$ is only slightly reduced from its value at 
zero--density, because of the slowness of the chiral symmetry restauration
(see the discussion in the previous section).

The mean field propagators of the scalar fields are thus
\be
D_0^{\sigma}(q_\mu)=1/(q_\mu^2-m^{*2}_{\tilde\sigma}+i\eta)
\ee
\be
D_0^{\chi}(q_\mu)=1/(q_\mu^2-M^{*2}_{\tilde\chi}+i\eta).
\ee

We define the scalar polarization propagator in the following way
\be
{\Pi}_s(q)=-i\int {d^{\,4}k\over (2\pi)^{4}}
Tr[G(k)G(k+q)],
\label{polprop}
\ee
where $G(k)$ is the single quark relativistic propagator {\it in
the medium} which can be decomposed in the Feynman propagator and a 
density dependent correction
\be
G(k)=(\gamma^\mu k_\mu+m_q)\left [{1\over k_\mu^2-m_q^2+i\eta}+
{i\pi\over E^*(k)}\delta(k_0-E^*(k))\theta(k_F-|\vec k|)\right ]
\equiv G_F(k)+G_D(k).
\ee
The effective quark mass is computed from mean field equations: 
$m_q=-g\bar\sigma/\bar\chi$, and $E^*(k)=\sqrt{m_q^2+k^2}$. For
a review of the formalism, see Reff.\cite{walecka,lim}.

In accordance with the mean field approximation, we neglect the vacuum
fluctuations effects, and we take only the density--dependent part of the
polarization propagator \cite{lim}. In such a way we avoid possible 
complications coming from the ill-defined zero density vacuum state.
An explicit expression for $\Pi_s$ can be found in Ref.\cite{lim}.

We now consider the propagators of the scalar fields, modified by the
polarization propagator insertion, i.e. the propagators in the RPA
approximation. The $\tilde\chi$ and the $\tilde\sigma$ propagators
mix up in the RPA approximation. They are the solution of the
following coupled equations
\begin{eqnarray}
D^{\sigma\sigma}&=&D_0^\sigma+D_0^\sigma \,g_{\tilde\sigma}\,\Pi_s
(g_{\tilde\sigma}D^{\sigma\sigma}+g_{\tilde\chi}D^{\chi\sigma})\nonumber
\\
D^{\chi\chi}&=&D_0^\chi+D_0^\chi \,g_{\tilde\chi}\,\Pi_s
(g_{\tilde\chi}D^{\chi\chi}+g_{\tilde\sigma}D^{\sigma\chi})\nonumber
\\
D^{\chi\sigma}&=&D_0^\chi \,g_{\tilde\chi}\,\Pi_s\,
(\,g_{\tilde\sigma}D^{\sigma\sigma}+g_{\tilde\chi}D^{\chi\sigma})\nonumber
\\
D^{\sigma\chi}&=&D_0^\sigma \,g_{\tilde\sigma}\,\Pi_s\,
(\,g_{\tilde\chi}D^{\chi\chi}+g_{\tilde\sigma}D^{\sigma\chi}).
\end{eqnarray}
In the previous equation, the dressed propagators correspond to various
situations in which the two mean field propagators $D_0^{\sigma}$
and $D_0^{\chi}$ mix among themselves {\it via} the polarization
propagator. These propagators can be decoupled, giving the following
RPA propagators
\begin{eqnarray}
D^{\sigma\sigma}&=&D_0^\sigma(1-g_{\tilde\chi}\,\Pi_s \,g_{\tilde\chi}D_0^\chi)
/\epsilon \nonumber
\\
D^{\chi\chi}&=&D_0^\chi(1-g_{\tilde\sigma}\,\Pi_s \,g_{\tilde\sigma}D_0^\sigma)
/\epsilon \nonumber
\\
D^{\chi\sigma}&=&D_0^\chi g_{\tilde\chi}\,\Pi_s \,g_{\tilde\sigma}D_0^\sigma
/\epsilon \nonumber
\\
D^{\sigma\chi}&=&D_0^\sigma g_{\tilde\sigma}\,\Pi_s \,g_{\tilde\chi}D_0\chi
/\epsilon \nonumber
\\
\epsilon&=&1-(g_{\tilde\chi}\Pi_s \,g_{\tilde\chi}D_0^\chi+
g_{\tilde\sigma}\Pi_s \,g_{\tilde\sigma}D_0^\sigma).
\end{eqnarray}
The collective excitations of the system correspond to the zeros of the
denominator $\epsilon(q_\mu)$. Instabilities are the collective states
at zero energy transferred \cite{lim}, i.e. the solutions of the equation
\be
\epsilon(q_0=0,\vec q\,)=0
\label{instab}
\ee
The solutions of this equation are shown in Fig.\ref{fig:INST}, as a function of
the transferred momentum and of the density of the system. In the region
enclosed by the line, corresponding to the solutions of eq. (\ref{instab}),
the system is instable; in particular, when these instabilities exist
for zero transferred momentum the system develops spontaneously
undamped density fluctuations. The range
of densities for which the system is found to be instable (at zero momentum
transferred) coincides with the range found studying the compressibility.

Concerning the interplay between chiral symmetry and confinement in the
development of the instabilities, we found again that chiral symmetry
plays a marginal r\^ole in the model: if one takes into account the
$\tilde\chi$ propagator only, the instability region shown in 
Fig.\ref{fig:INST} is almost unchanged.

\section{two--body correlations: gap equation}

We will now study two--body correlations, using the formalism
of the gap equation \cite{fetter}. The gap is the difference
between the energy of the incorrelated pair and the energy of the
correlated pair. It is thus positive if the potential is attractive.
We will use as the residual interaction the one arising from the exchange
of a $\tilde\chi$ or a $\tilde\sigma$. 
%We want to check if the region
%of densities in which the system is instable corresponds to the region
%where a gap develops. In such a case we can interprete the instability
%of the system as resulting from the increase of the correlations between
%couples of quarks. 
Of course we are able to take into account only two--body
correlations, and three body--correlations can be even more important. 

The gap equation reads \cite{fetter}
\be
\Delta_k={1\over 2}\sum_{\bf k'} <{\bf k},-{\bf k}|V|{\bf k'},-{\bf k'}>
{\Delta_{k'}\over (\Delta_{k'}^2+\xi_{k'}^2)^{1/2}}
\label{gap}
\ee
Here we are following the convenction \cite{fetter} that an attractive
potential is positive. The potential $V$ is evaluated between 
incorrelated states, described by plane waves.
$\xi_k$ is the single--particle energy, mesured relative to the chemical
potential $\mu=\sqrt{k_F^2+m_q^2}$. We will not use the relativistic reduction
of $\xi_k$, because the relativistic corrections are not totally negligible.
Therefore $\xi_k=\sqrt{k^2+m_q^2}-\sqrt{k_F^2+m_q^2}$, where $m_q$ is
the quark mass as computed in the mean field approximation.

As we have already said, the potential $V$ arises from the exchange
of the scalar fields' fluctuations $\tilde\chi$ and $\tilde\sigma$.
The masses of this fluctuations and the couplings between the 
fluctuations and the quarks' fields, have been obtained in the
previous section, expanding the lagrangian around the mean field
approximation. Of course this expansion, which was sufficient to study
the instabilities of the system, is in general quite a crude approximation.
The results obtained for the gap are thus only the first step in the
study of the difficult problem of the clusterization.

The potential $V$ is a Yukawa potential, but an extra factor
$m_q^2/E^*(k)E^*(k')$ has to be included ($E*(k)=\sqrt{k^2+m_q^2}$), 
because of the choosen normalization of the quarks' spinors.

Following Ref.\cite{fetter}, since the resulting gap $\Delta_k$ is much
smaller than the Fermi energy, the integrand appearing in eq.(\ref{gap})
is sharply peaked near $\xi=0$, and then 
$\Delta_k\simeq\Delta_{k_F}\equiv\Delta$, i.e. the gap is almost independent
on the momentum.

We show in Fig.\ref{fig:INST} the resulting gap. It shows a strong dependence on the
density, and for densities of the order of $\rho_{eq}$ is already
totally negligible.

The gap equation can be solved analytically for small value of the gap.
The result is
\be
\Delta\approx 8{k_F^2\over 2 m_q}{\rm exp}
(-{\pi k_F^2/ 2 m_q\over <k_F|V|\phi_{k_F}>}),
\ee
where the matrix element of the potential is 
\be
<k_F|V|\phi_{k_F}>\equiv k_F\int_0^\infty {\rm sin}k_FxV(x){\rm sin}k_Fx
{\rm d}x.
\ee
Since our residual interaction $V$ is always attractive, it doesn't exist a
critical density at which the gap is exactly zero, and this value
is reached only asymptotically.

\section{conclusions}

In this paper we have studied quark matter, using the non--perturbative
tool of the CDM
model. Let us summarize our main results:

-- of all the considered versions of the model, only one gives sensible
results, i.e. the one in which confinement is imposed in the smoothest way.
The other versions of the model give an exceedingly low energy per baryon
number for the quark matter.

-- the SM (p=1) version gives an EOS for the quark matter 
which is almost identical to the
EOS of nuclear matter as computed using the Walecka model, for the
range of densities $\rho_{eq}\le\rho\le 2\rho_{eq}$. In this range, 
the difference in the energy per baryon number between the nuclear matter
and the quark matter is very small. Taking into account 
the theoretical incertitude
in the fixing of the parameters (Sec.2B), this energy difference is of
the order of 20 MeV. At densities smaller than $\rho_{eq}$ the energy
difference rapidly increases, and for densities higher than $2\rho_{eq}$
the quark matter is the energetically most favourable state. An important
point is that the minimum for the EOS of quark matter is at a density
of the order of $\rho_{eq}$, and this result does not depend on the
fine tuning of the parameters.

-- the mass of the quarks remains big (of the order of 100 MeV) till
high densities, much higher than the density at which quark matter
becomes the ground state. The deconfinement phase transition and 
the chiral symmetry restauration arise at totally different densities.

-- the quark matter (in SM and p=1) becomes unstable at low densities,
of the order of $\rho_{eq}/2$. The instability can be obtained both from
the study of the compressibility (where the compressibility becomes
negative the system is unstable) and from the study of the collective
states at zero energy transfer. The two method give the same critical
density.

-- the process of clusterization can be studied considering the correlations
between the particles, beyond the mean field approximation. The gap
we obtained seems rather small. We have to bear in mind that we have
oversimplified the problem, by linearizing the residual interaction
(and thus getting an approximate propagator for $\tilde\chi$, good only
for small fluctuations), and by considering only two body correlations,
where the three body ones are probably the most relevant.

To conclude, we would like to consider three possible applications of the model.

-- Cooling of neutron stars
(see C.J.Pethick in \cite{brown}). 

A mechanism called URCA has been invoked to explain
the rapid cooling of neutron stars. This mechanism proceeds via the exchange
of electrons between neutrons and protons, which cool down emitting
neutrinos and antineutrinos:
\be
n\rightarrow p+e^{-}+\bar\nu_e
\ee
\be
p+e^{-}\rightarrow n+\nu_e
\ee
A minimal fraction of protons is required, in order to fulfil momentum
and energy conservation. This critical fraction is of the order of 1/9. Using
traditional nuclear physics models to compute the protons' fraction,
one gets numbers slightly smaller than the critical one, and the
URCA mechanism cannot start.

Another possibility is to invoke the presence of quark matter in the core
of the star, and to consider reactions similar to the one previously
described, but with the electron now exchanged between up and down quarks.
In this case the problem is that, considering quark matter as described by
the MIT bag model, quarks are massless and the phase space is thus zero.
Therefore one would need a massive quark matter phase, and the possibility
of reaching this phase at the density of the core of neutron
stars, tipically of the order of $5\rho_{eq}$. This situation is actually
the one described by the SM (p=1) version of the CDM. URCA mechanism
should therefore be possible, and with an high luminosity, too.

-- Energy released in supernova explosion.

Using a traditional nuclear physics approach, the energy released in 
supernova explosion is generally too small. A softer EOS could solve
the problem, but if one uses e.g. the MIT bag to study matter at high
density, the deconfinement phase transition is reached at 
densities larger than the one presumably reached in the collapse of the
star. The EOS for matter at high density as computed in the 
CDM, is softer than the EOS of nuclear matter, and presumably a similar
result will be obtained when computing neutron matter. Furthermore the
softening starts at densities of the order of $2\rho_{eq}$.

-- EMC effect and swelling of the nucleon.

To conclude let us consider the problem of the possible swelling of 
the nucleons embedded in a nucleus. If one considers,e.g. electron--scattering
on heavy nuclei, one realizes that the swelling is a sensible mechanism,
but it must be of the order of some $5\%$ in order to be realistic.
The real problem is thus not the one to obtain a swelling, but to 
obtain a not too big effect. In other words, the nucleons have not to
dissolve when embedded in a nucleus. Since the minimum of the EOS of
the quark matter is for a density near $\rho_{eq}$, only in the center
of heavy nuclei some swelling mechanism can appear. The exact amount
of swelling depends on the precise difference in energy between the 
quark matter and the nuclear matter at densities $\sim\rho_{eq}$, and
is beyond the possibility of the present calculation.

We are now carrying out researches in all the directions previuosly outlined.
%%%%%%%%%%%%%%%%%%%%%%%%%%%%%%%%% ACKNOWLEDGMENTS
%\acknowledgments 

%%%%%%%%%%%%%%%%%%%%%%%%%%%%%%%%% APPENDIX
\appendix
\section*{}

In this appendix we discuss the behaviour of the $\chi$ field
in the DM version of the CDM, for a two nucleon system, as a function
of their distance $d$, which is related to the density of the system
$\rho\propto d^{-3}$.
We will show that a critical internucleon distance 
exists, at which the $\chi$ field has a discontinuous behaviour.

For simplicity we will omit the chiral fields, the pion and the sigma,
from our discussion, considering the following Lagrangian:

\be
{\cal L} = i\bar \psi \gamma^{\mu}\partial_{\mu} \psi
       -{g\over \chi} \, \bar \psi f_\pi \psi       
         +{1\over 2}{\left(\partial_\mu\chi\right)}^2
                       -U \left(\chi\right)\, ,
\label{eq:ina1}
\ee
where the potential for $\chi$ is the one given in eq.(3) and shown
in Fig.5a. We have labelled by $\chi_m$ the relative minimum and
by $\chi_1$ the other value of $\chi$ for which 
$U(\chi_1)=U(\chi_m)=M^4\eta^4$.

We consider a system made of two clusters of three quarks each, with
an intercluster distance $d$.
In the mean field approximation 
the total energy of the two--nucleon system is given by
\be
E_2=6 \epsilon_{q} + T_\chi + E_\chi
\label{eq:ina2}
\ee
where $\epsilon_q$ is the energy of the single quark, 
$T_\chi$ is the energy of the $\chi$
field associated with its spatial fluctuations \footnote{This energy
is not the kinetic energy of the $\chi$ field, which is zero, because
$\chi$ is a scalar field and it is assumed to be time independent.} and
$E_\chi$ is the energy coming from the potential $U(\chi)$.

When the intercluster distance $d$ is large, the $\chi$ field interpolates
between a number slightly bigger than $\chi_m$ and a small number, smaller
than $\chi_1$, in the internucleon space (see Fig.5b, the solid and dotted
line).

When the distance $d$ is reduced, the value of the scalar field in the
internucleon region increases. We compare now two possible solutions:
one in which the $\chi$ field has a minimum value in the internucleon
region, equal to $\chi_1$, and a second in which the $\chi$ field
remains almost constant in the internucleon region (solid and dashed line
in Fig.5b). The second solution has a smaller energy
than the first one. In fact: 
$\epsilon_q$ is smaller, 
because the quarks move in a single big well, instead of moving in a 
double well;
$T_\chi$ is smaller because $\chi$ is fluctuating less;
$E_\chi$ is smaller because $\chi$ is not moving (in the internucleon region)
through the relative maximum of $U(\chi)$.
Since the second solution has a smaller energy, 
the value of $\chi$ in the internucleon region will not
smoothly increse, as the distance $d$ is decreased, but will jump,
at a certain critical distance, 
from the behaviour described by the solid and dotted
line in Fig.5b to the one described by the solid and dashed line.
The internucleon value of $\chi$ can be assumed as an order parameter
that undergo a discontinuous change as the density is increased, thus
the transition is first order.

%
%%%%%%%%%%%%%%%%%%%%%%%%%%%%%%%%% REFERENCES LIST
%

%
%%%%%%%%%%%%%%%%%%%%%%%%%%%%%%%%% FIGURE CAPTION

\begin{figure}
\centerline{\hbox{
\psfig{figure=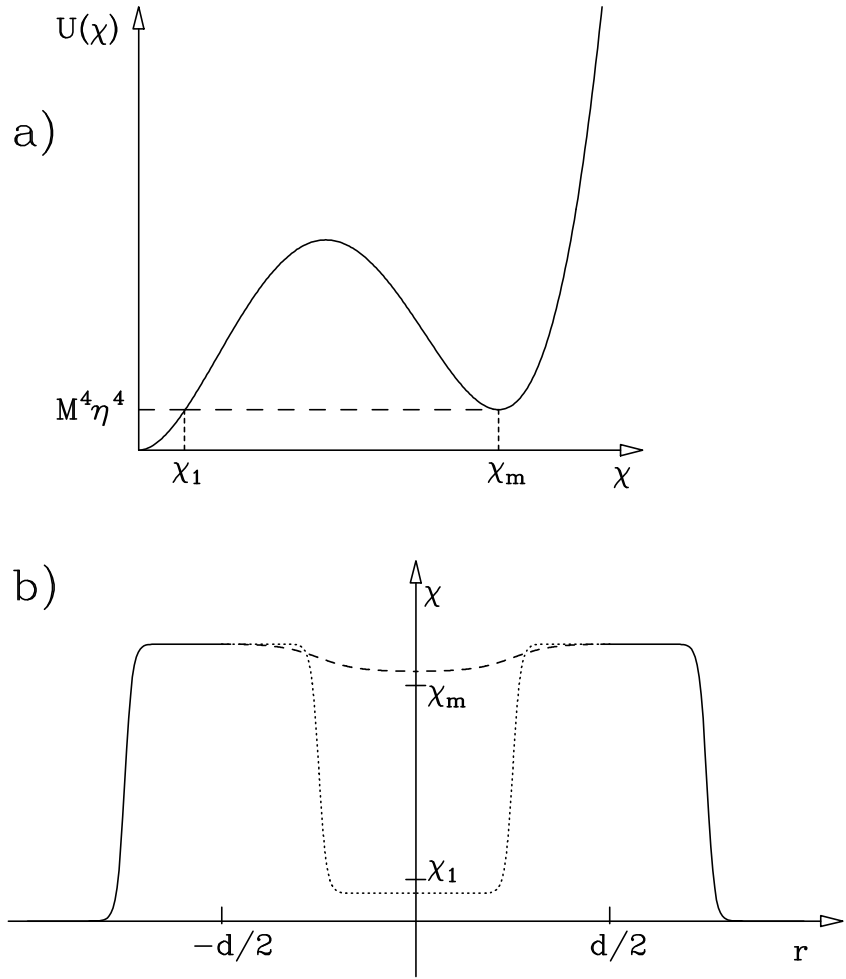,angle=0,width=12cm}
}}
\caption{
Behaviour of $\chi$ field in the DM version of the CDM.
Potential $U(\chi)$ (a) and shape of $\chi$ field for internucleonic
distance $d$ 
.}
\label{fig:APP}
\end{figure}

\end{document}